\documentclass[12pt]{iopart}
\usepackage[final]{graphicx}
\usepackage{iopams}
\usepackage{bm}
\usepackage{cite}
\usepackage{epsfig,color}
\usepackage{xcolor}
\usepackage{hyperref}
\hypersetup{
    colorlinks = true
}
\usepackage{microtype}

\usepackage{float,siunitx}
\usepackage[caption = false]{subfig}

\usepackage[greek,english]{babel}
\usepackage{thumbpdf,enumerate}
\usepackage{booktabs}
\usepackage{sidecap}
\usepackage[scaled=.8]{couriers}
\usepackage{multirow}
\usepackage{placeins}
\usepackage{relsize}
\usepackage{epigraph}
\usepackage{gensymb}
\usepackage{longtable}
\usepackage{soul}
\usepackage{ulem} 
\usepackage{acronym}

\newcommand{\ket}[1]{\left\vert{#1}\right\rangle}

\begin{document}

\title{Enhanced detection techniques of Orbital Angular Momentum states in the classical and quantum regimes}

\author{Alessia Suprano$^{1}$, Danilo Zia$^{1}$, Emanuele Polino$^{1}$, Taira Giordani$^{1}$, 
Luca  Innocenti$^{2,3,4}$, Mauro Paternostro$^{3}$, Alessandro Ferraro$^{3}$, Nicol\`o Spagnolo$^{1}$ and Fabio Sciarrino$^{1}$ }

\address{$^{1}$Dipartimento di Fisica, Sapienza Universit\`{a} di Roma, Piazzale Aldo Moro 5, I-00185 Roma, Italy}

\address{$^2$Department of Optics, Palack\'{y} University, 17. Listopadu 12, 771 46 Olomouc, Czech Republic}
\address{$^3$Centre for Theoretical Atomic, Molecular, and Optical Physics,
School of Mathematics and Physics, Queen's University Belfast, BT7 1NN Belfast, United Kingdom}
\address{$^4$Università degli Studi di Palermo, Dipartimento di Fisica e Chimica – Emilio Segrè, via Archirafi 36, I-90123 Palermo, Italy}

\ead{fabio.sciarrino@uniroma1.it}
\vspace{10pt}

%
%
%
%
%

\begin{abstract}
The Orbital Angular Momentum (OAM) of light has been at the center of several classical and quantum applications for imaging, information processing and communication. However, the complex structure inherent in OAM states makes their detection and classification nontrivial in many circumstances. Most of the current detection schemes are based on models of the OAM states built upon the use of
Laguerre-Gauss modes.
However, this may not in general be sufficient to capture full information on the generated states. In this paper, we go beyond the Laguerre-Gauss assumption, and employ Hypergeometric-Gaussian modes as the basis states of a refined model that can be used -- in certain scenarios -- to better tailor OAM detection techniques.
We show that enhanced performances in OAM detection are obtained for holographic projection via spatial light modulators in combination with single-mode fibers, and for classification techniques based on a machine learning approach. 
Furthermore, a three-fold enhancement in the single-mode fiber coupling efficiency is obtained for the holographic technique, when using the Hypergeometric-Gaussian model with respect to the Laguerre-Gauss one.
This improvement provides a significant boost in the overall efficiency of OAM-encoded single-photon detection systems.
Given that most of the experimental works using OAM states are effectively based on the generation of Hypergeometric-Gauss modes, 
our findings thus represent a relevant addition to experimental toolboxes for OAM-based protocols in quantum communication, cryptography and simulation.

\end{abstract}

\maketitle

\section{Introduction}

Light supports the Orbital Angular Momentum (OAM) degree of freedom, whose states are encoded in a structured spatial transverse amplitude profile ~\cite{rubinszteindunlop2016roadmap,padgett2004light,erhard2018twisted}.
Such wavefronts are also frequently referred to as \textit{twisted light}, due to their characteristic helicoidal phase profile~\cite{erhard2018twisted}.
In the classical domain, the OAM of light finds several applications, including optical trapping and laser micro-machining~\cite{zhan2009cylindrical}, micro-manipulation~\cite{dholakia2008optical}, optical communication~\cite{gibson2004free,willner2015optical}, microscopy~\cite{furhapter2005spiral,tamburini2006overcoming}, sensing~\cite{lavery2013detection}, and high-density optical information encoding~\cite{wang2012terabit,bozinovic2013terabitscale,willner2015optical}.
OAM is also a useful quantum resource 
and can be exploited as an alphabet to encode high dimensional quantum states~\cite{erhard2018twisted}.
More generally, OAM states represent an invaluable tool in several quantum information protocols, including quantum communication ~\cite{Wang2015,Mirhosseini_2015,krenn2015twisted,Malik2016,malik2012influence,baghdady2016multi,Sit17,Cozzolino2019_fiber,wang2016advances}, quantum simulation \cite{cardano2016statistical,cardano_zak_2017,Buluta2009}, quantum computation and error correction protocols~\cite{bartlett2002quantum,ralph2007efficient,Lanyon2009,Michael2016}.
A particularly interesting class of states carrying nontrivial OAM is obtained coupling spatial and polarization degrees of freedom of light.
Such states, often referred to as \textit{vector vortex beams} (VVBs), have been used in the contexts of optical trapping~\cite{Cardano2015Rev, rubinszteindunlop2016roadmap}, metrology~\cite{dambrosio2012complete,fickler2012quantum,polino2020photonic,kozawa2018superresolution,suprano2020propagation,berg2015classically}, and communication~\cite{Sit17,Cozzolino_rev,Cozzolino2019_fiber,cozzolino2019air,dambrosio2012complete,vallone_qkd_2014}.

Several platforms have been proposed to engineer and manipulate OAM states in photonics~\cite{rubinszteindunlop2016roadmap}, including pitch-fork
gratings~\cite{bazhenov1992screw}, computer-generated holograms with spatial light modulators (SLM)~\cite{kirk1971phase,heckenberg1992generation,bolduc2013holo,forbes2016creation,arrizon2007pixelated}, $q$-plates (QPs)~\cite{marrucci2006optical, piccirillo2013orbital}, cylindrical lenses~\cite{beijersbergen1993astigmatic,shen2021creation}, spiral phase plates~\cite{beijersbergen1994helical}, ring resonators~\cite{cai2012integrated}, metasurfaces~\cite{jin2016generation,devlin2017arbitrary,deng2018diatomic},  lasers \cite{naidoo2016controlled,sroor2020high,forbes2019structured,shen2020structured} and interferometric or refractive devices \cite{padgett1996experiment,berkhout2008method}.
Generation of VVBs was also recently demonstrated in integrated photonic chips~\cite{chen2020vector}.

Any application of OAM states crucially relies on the capability to detect these states accurately.
However, reconstructing high-dimensional OAM states is in general challenging due to the large dimension of such problem, and the complexity of the associated spatial profiles.
This is particularly the case in the quantum domain, where the full characterization of OAM states requires quantum state tomography, which is very demanding for high-dimensional states~\cite{thew2002qudit}.
Known techniques to characterize OAM states include demultiplexing based on multi-plane light conversion~\cite{labroille2014efficient}, holographic and optical geometric transformation-based sorters~\cite{kaiser2009complete,genevet2012holographic,Berkhout2010_oamSorter,mirhosseini2013efficient,lavery2012refractive,o2012near,slussarenko2010polarizing,Leach2002_oamSorter,d2017measuring},  metasurfaces~\cite{jin2016generation,devlin2017arbitrary,deng2018diatomic,gao2018nonlinear,guo2021spin}, holograms imprinting phase patterns (\textit{e.g.} from SLMs) followed by single-mode fibers~\cite{mair2001entanglement,forbes2016creation,Qassim2014,giordani_2018,bouchard2018measuring}, hybrid approaches \cite{nape2020enhancing}, time-based multiplexing \cite{bierdz2013high,bierdz2011compact,karimi2012time}, lenses \cite{komal2021polarization}, techniques using Doppler frequency shifts \cite{courtial1998rotational,vasnetsov2003observation,zhou2017orbital}, interferometric, refractive or diffractive schemes \cite{padgett1996experiment,Leach2002_oamSorter,berkhout2008method,slussarenko2010polarizing,ferreira2011fraunhofer,lavery2012refractive,mazilu2012simultaneous,mourka2011visualization,ariyawansa2021amplitude} and weak measurements~\cite{Mehul2014_oamSorter}.
Machine learning (ML) methods have also recently proved valuable in the context of the reconstruction of the properties of structured light. In particular, supervised and unsupervised learning techniques were used to classify OAM states propagating through free-space~\cite{doster2017machine, Park_18,na2021deep}  and through turbulent environments~\cite{Krenn13648, krenn_2014,Lohani_18,liu2019deep,li2018joint, NEARY2020126058,Xie:15,lohani2018turbulence,bhusal2020spatial,silva2020machine,zhan2021experimental}, as well as to classify and reconstruct VVB states~\cite{giordani2020machine,zhai2020turbulence}.



\begin{figure*}[ht]
\includegraphics[width=\textwidth]
   {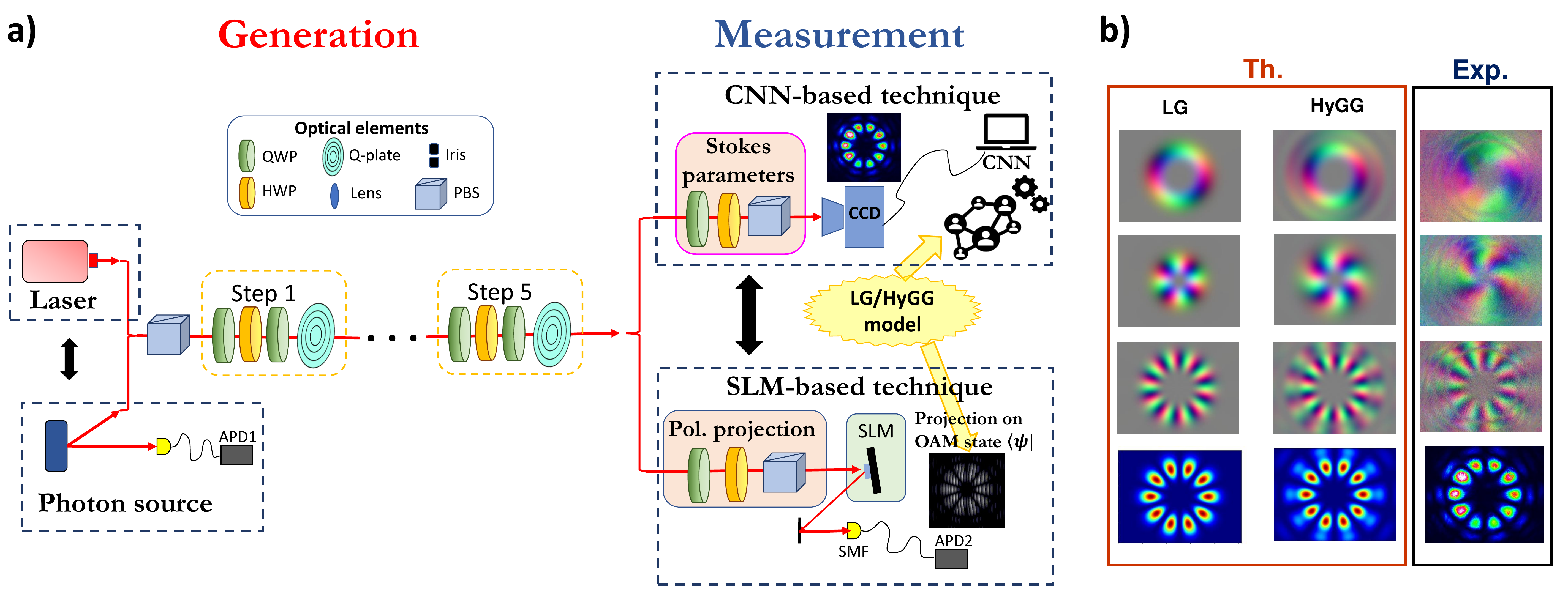}
    \caption{\textbf{Experimental generation and detection of OAM-based states.} a) The employed experimental platform is based on a five-step quantum walk to generate OAM and VVB states with both single-photon inputs and classical laser light. Each step is composed of a set of two quarter waveplates (QWP) interspersed by a half waveplate (HWP) and a QP. The states resulting from these arrangements are measured using two different detection apparatuses. In the classical domain, the detection system is composed of the polarization analyser and a Charge-Coupled Device camera (CCD). This arrangement measures the spatial distribution of the input beam and sends the acquired images to a computer that classifies the states using a suitably trained Convolutional Neural Network (CNN). In the single-photon domain, the OAM measurement stage consists of a polarization analyser and an SLM followed by the spatial filter provided by a single-mode fiber (SMF) and connected to a single-photon detector (APD). b) Example of experimentally measured patterns generated by the quantum walk platform (right), and the corresponding theoretical predictions obtained via an LG and a HyGG model, respectively. The first three rows represent the colored maps of three different VVBs corresponding to $\{m_1,m_2\}=[\{3,5\},\{-1,3\},\{5,-5\}]$. Each color in the map is associated with a different direction of the polarization and the distribution of the colors depends on the value of $m_1$ and $m_2$. Instead, in the last row is reported the pattern distribution associated with OAM state obtained as the balanced superposition of $m_1=5$ and $m_2=-5$.}
    \label{fig:app}
\end{figure*}

The efficiency of detection techniques based on the cascade of SLMs and single-mode fibers \cite{cardano2015quantum} or ML algorithms for the classification of experimental states based on the pattern distribution of theoretical modes \cite{doster2017machine,Park_18, giordani2020machine}, depends crucially on the provision of accurate models of states  produced by the experimental apparatus in use.
Most of the proposed and tested methods 
consider the experimental state described by the Laguerre-Gauss (LG) model.
However, this model might not fully capture the structure of the optical modes generated by any given experimental apparatus, which would limit the accuracy of the detection scheme. 
The performance of such detection schemes would thus be improved by the use of more accurate models for the incoming light. Indeed, the output beams from devices like QPs and SLMs are usually modeled with LG modes, which thus neglect any effect due to diffraction. 
This approximation is valid in the pupil plane under the assumption of a thin device.
On the contrary, in a generic transverse plane the radial index number, which characterizes the transverse spatial distribution of OAM states, 
does not have a fixed value~\cite{karimi2009light,sephton2016revealing}, although the output beam has a well-defined OAM value. 
In this scenario, the beams that are generated by the considered device \cite{karimi2009generation,karimi2009light,shu2016propagation,sephton2016revealing} can be describes more accurately by a model based on Bessel \cite{durnin1987diffraction,kotlyar2014asymmetric} and Hypergeometric-Gaussian (HyGG) functions~\cite{karimi_07}, whose propagation has also been studied in turbulent enviroment \cite{eyyubouglu2012hypergeometric,zhu2015effects,cheng2016channel,bian2018propagation}. 

In this paper, we develop an improved approach to model the state of OAM modes which provides enhanced characterization performances. We employ HyGG modes to model experimentally OAM states engineered via a quantum-walk-based platform composed of a series of QPs in a cascaded configuration. We test our platform both in the classical and quantum regimes.
To assess the improved measurement performances brought forward by adopting our HyGG-based model, we use it to create computer-generated holograms exploited in the SLM-based measurement. We observe higher state fidelities when coupling the resulting output modes to single-mode fibers. Consequently, the exploitation of the refined model allows us to extract more information about the experimental state regarding both the azimuthal and the radial distribution. Furthermore, this result has been obtained by keeping fixed the number of holograms used to perform the projection measurement, without using distinct hologram for each Laguerre-Gaussian mode with different radial indexes. These are associated with substantially higher coupling efficiencies.
We also show that states generated using the HyGG model, when used to train a CNN, provide enhanced learning capabilities, 
thus allowing to better predict experimental images.
These results highlight the importance of using an accurate model of the incoming beams to optimize the detection process and could be adopted for different techniques that need an accurate model of the generated state.

The remainder of this paper is organized as follows. In Sec.~\ref{generazione}, we provide a brief description of the experimental apparatus of the generation and detection of the OAM states here at hand. Sec.~\ref{modello} is dedicated to the illustration of the modelling used to describe the OAM state produced by the experimental platform. Sec.~\ref{ologramma} makes use of our HyGG model to capture the experimental features through computer-generated holograms. In Sec.~\ref{ML}, we put in place a ML-based classification approach. Finally, in Sec.~\ref{conc} we draw our conclusions. A series of technical details of our analysis is deferred to the Appendices.

\section{Generation and detection of arbitrary OAM states}
\label{generazione}

OAM of light is encoded in the helicoidal transverse spatial wavefront of an optical beam.
More specifically, OAM-endowed modes are characterized by a phase dependence of the form $e^{im\phi}$, with $\phi$ the azimuthal angle in the transverse plane with respect to the propagation direction.
Coupling OAM modes with non-uniform polarization distributions produced the so-called VVBs, which are superpositions of two different helicoidal fields $\vec{E}_{m_1}$ and $\vec{E}_{m_2}$ with OAM parameters $m_1$ and $m_2$, and orthogonal polarizations $\vec{e}_L$ and $\vec{e}_R$. The resulting field would read $\vec{E}_{m_1,m_2}=\vec{e}_L\cos(\theta/2)\vec{E}_{m_1}+e^{i\beta}\vec{e}_R\sin(\theta/2)\vec{E}_{m_2}$ with  $\theta \in [0,\pi]$ and $\beta \in [0,2 \pi]$.

To engineer OAM states characterized by OAM numbers in $\{\pm1,\pm3,\pm5\}$, we exploit an experimental platform implementing a scheme based on one-dimensional discrete-time quantum walks (QWs) exploiting both the OAM and polarization degrees of freedom~\cite{giordani_2018}.
In particular, the walker (coin) state of the QW is encoded in the OAM (polarization) degree of freedom. The QPs change the OAM state conditionally to the polarization, thus implementing the QW step operation.
The implementation of the QW is performed through a cascade of five QPs interspaced by a set of waveplates (WPs) (Fig. \ref{fig:app} a). A QP is a slab of a birefringent material characterized by a non-uniform transverse optical axis pattern with a topological charge equal to $q$. The action of this device is to add a phase term $e^{\pm i 2q \phi}$ to the incoming beam depending on the polarization state, thus resulting in a variation of the OAM value by $\pm 2q$~\cite{Marrucci2011Rev}. At each step of the protocol, a sequence of three WPs and a QP enables the engineering of arbitrary modes for increasingly larger OAM values. This platform is advantageous with respect to more usual methods for the generation of OAM states based on the use of SLMs in light of its flexibility. In particular, VVB states can be engineered naturally employing QPs~\cite{Cardano:12} without the need for interferometric architectures~\cite{Ndagano:18, Maurer2007}. 

We employ two different strategies to classify engineered optical states, and test these approaches in two distinct regimes, as shown in Fig.~\ref{fig:app} (a).
Although both hologram-based and ML-based approaches can be used in quantum and classical regimes,  we perform a complete investigation exploiting the holographic technique for single-photon states~\cite{bolduc2013holo,forbes2016creation}, and a CNN for detecting classical laser states. 
As a first step, a projective measurement is performed in the horizontal polarization basis via a polarizing beamsplitters (PBS). Subsequently, a SLM is employed to convert a given OAM mode in the fundamental Gaussian mode, which is then coupled into a single-mode fiber. This operation corresponds to a projective measurement of the walker state encoded in a chosen arbitrary superposition of OAM components.
The similarity between the target state and the experimentally generated walker state is quantified through the quantum state fidelity, which is measured by performing different projection of the input mode onto an orthonormal basis that contains the given target state.

On the other hand, we use CNNs to classify the generated VVB states.
As input for the classification algorithm  we use the intensity profiles as measured by a CCD camera. In order to capture the information stored in the \textit{polarization} patterns, we measure the intensity after projection along each polarization direction. 
 A standard approach to do this is to choose the three mutually unbiased polarization bases, $b_1=\{H,V\}$, $b_2=\{D,A\}$, and $b_3=\{L,R\}$, measure the intensity profiles $I_{b_{i,j}}$ associated with each of them, and then compute the so-called \textit{Stokes parameters} $S_i=(I_{b_{i,1}}-I_{b_{i,2}})/(I_{b_{i,1}}+I_{b_{i,2}})$.
 The Stokes parameters characterize the polarization state at each point of the transverse profile.
To visualize the intensity profiles thus obtained, we represent the values of the Stokes parameters using an RGB encoding, with the value of each $S_i$ mapped into the intensity of one of the primary colours red, green, and blue%
~\cite{giordani2020machine}.
The processed images are then analyzed through a suitably trained CNN to perform recognition of the incoming mode.

\section{Modeling the output states of the quantum walk platform}
\label{modello}

Modes carrying OAM values can be described using specific Helmholtz equation solutions in the paraxial approximation.
In particular, the most commonly employed are the LG modes. They are characterized by two integer indexes $(p,m)$, where the former is associated with the radial structure of the beam, while the second describes the azimuthal phase structure. 

As outlined above, the platform used to engineer arbitrary OAM modes and VVB states employs a cascade of five QPs, interspersed with WPs to modulate the polarization state. The action of a QP is commonly associated only to a modification of the azimuthal index $(m)$ of the input LG mode ($\text{LG}_{p,m}$), while keeping the radial index $(p)$ unchanged. 
Thus, the action of a QP is described as \cite{Marrucci2011Rev}
    \begin{eqnarray}
    &\text{LG}_{p,m}\vec{e}_{L} \longmapsto \cos\frac{\delta}{2}\;\text{LG}_{p,m}\vec{e}_{L} + i e^{i 2\alpha_{0}}\sin\frac{\delta}{2}\;\text{LG}_{p,m+2q}\vec{e}_{R}, \nonumber \\
    &\text{LG}_{p,m}\vec{e}_{R} \longmapsto \cos\frac{\delta}{2}\;\text{LG}_{p,m}\vec{e}_{R} + i e^{-i 2\alpha_{0}}\sin\frac{\delta}{2}\;\text{LG}_{p,m-2q}\vec{e}_{L}.
    \label{Eq.QP}
\end{eqnarray}
Here $\delta \in [0,\pi]$ is the QP uniform birefringent phase retardation, while $\alpha_{0}$ is the initial angle between the optical axis and the $x$ axis on the device plane. In our discussion we consider only an optimal tuning of the QP, obtained for $\delta = \pi$.

 Such approximation of the QP action is valid only in the pupil plane of the device, where the diffraction effect can be neglected. Conversely, a variation of the radial index has to be considered in a general scenario. To consider such effect and go beyond the approximation described by ~(\ref{Eq.QP}) we derived the beam after propagation through each QP by solving the Fresnel's integral
\begin{eqnarray}
    E_{\text{out}}(x,y,z) &= -\frac{e^{-i k(z-\tilde{z})}}{i \lambda (z-\tilde{z})}\int_{-\infty}^{\infty}\int_{-\infty}^{\infty}d\tilde{x}d\tilde{y}\;E_{\text{in}}(\tilde{x},\tilde{y},\tilde{z})\times \nonumber \\ & \times\exp\left\{-i  \frac{k\, d^2_\perp}{2\left(z-\tilde{z}\right)} \pm 2i (q\phi + \alpha_{0})\right\}
    \label{Eq.Fresnel}
\end{eqnarray}
with $d_\perp=\left[\left(x-\tilde{x}\right)^{2}+\left(y-\tilde{y}\right)^{2}\right]^{1/2}$.
Here $\tilde{z}$ is the position of the QP, $(\tilde{x}, \tilde{y})$ are the coordinates in the transverse plane of the device and $\phi = \arctan(\tilde{y}/\tilde{x})$ is the azimuthal angle. In particular, for the QW described above the integral in ~\ref{Eq.Fresnel} has to be solved by considering a Gaussian input mode for the first QP. Then, the input field in the $n$-th QP $E_{\text{in}}(\tilde{x},\tilde{y},\tilde{z})$ is the output beam of the $(n-1)$-th step. The result of the integration for the first QP is a HyGG mode ($\text{HyGG}_{p,m}$) \cite{karimi2009light,shu2016propagation}, which is characterized by the azimuthal number $m=\pm 2q$ and a real parameter $p \geq -\vert m \vert$. By exploiting the following relation \cite{karimi_07}
\begin{equation}
\text{HyGG}_{p,m} = \sum_{k=0}^{\infty} A_{p,k}\text{LG}_{k,m},
\label{eq:HyGG_as_LG}
\end{equation}
where 
\begin{equation}
    A_{p,k} = \sqrt{\frac{(k + |m|)!}{k!\;\Gamma(p +|m|+1)}}\frac{\Gamma(k-p/2)\Gamma(p/2+|m|+1)}{\Gamma(-p/2)\Gamma(k+|m|+1)},
\end{equation}
the output mode can be written as an infinite superposition of LG modes (here $\Gamma(x)$ is the Gamma function).

At variance with previous approaches, which consider only the first term of the superposition ($k=0$), we include all terms up to order $k=3$ as input of the second QP. This choice allows us to reach a higher overlap between the theoretical and experimental images without substantially increasing the computational cost. After solving the integral in ~\ref{Eq.Fresnel} for each considered LG mode, the output beam from the second QP is described by a finite superposition of HyGG modes. Exploiting again ~\ref{eq:HyGG_as_LG}, the procedure is repeated for all the QPs of the setup to obtain a final description of the output beam.
Consequently, the output state engineered via the cascaded platform is expressed as a superposition of HyGG modes with different radial indices but with the same azimuthal index of the target state (more details are reported in the Appendix A).
Therefore, such theoretical model goes beyond the LG assumption, and provides  a more accurate description of the beam propagation inside the quantum walk platform (an example is shown in Fig. \ref{fig:app} b). In the following sections, we 
exploit this refined model to reach enhanced performances in OAM detection techniques using both holographic projection and machine learning-based approaches.

\begin{figure}[hbt]
    \centering
    \includegraphics[width=0.7\columnwidth]{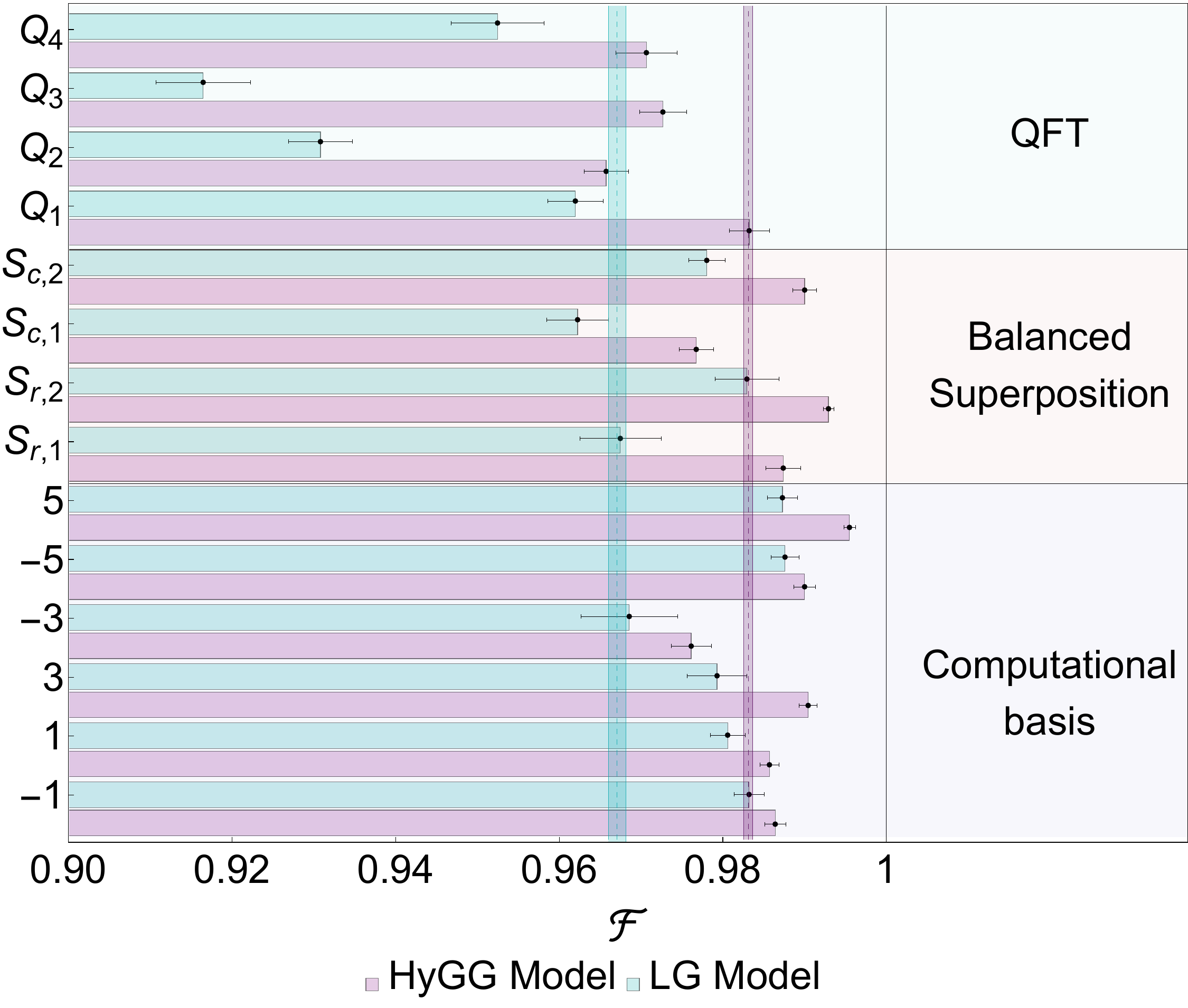}
    \caption{\textbf{Experimental quantum state fidelity.} Summary of quantum state fidelities obtained by performing the measurements of the 14 engineered states with the holograms based on the HyGG model (purple bars) or the LG one (cyan bar). We reported the comparison for different classes of state: elements of the computational basis, superposition of large OAM states (Balanced Superposition) and elements of the Fourier basis (QFT). The purple and cyan lines that represent the mean values are $0.9831 \pm 0.0005$ and $0.9671 \pm 0.0010$, respectively. }
    \label{fig:Fid}
\end{figure}

\begin{figure}[htb]
    \centering
    \includegraphics[width=0.7\columnwidth]{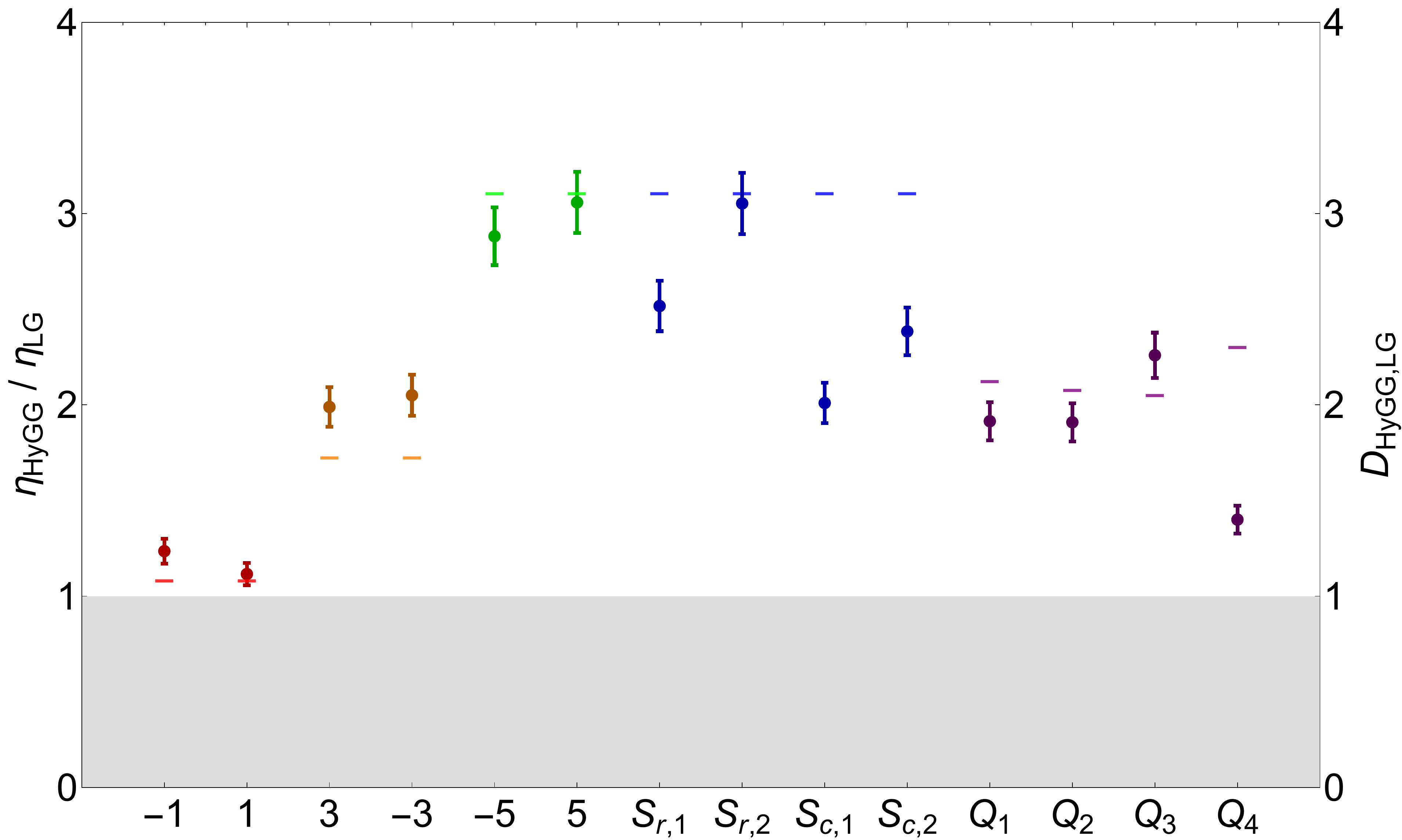}
    \caption{\textbf{Coupling Efficiency.} The points with the experimental error bars represent the ratio between the coupling efficiency associated to the holograms programmed using the HyGG model and that one obtained with the LG model. The ratio is always greater than 1, confirming the coupling improvement.
    Moreover, the experimental values are compared with the quantity $D=\frac{|\langle \Phi_{\text{HyGG}} \vert \Phi_{\text{exp}} \rangle|^2}{|\langle \Phi_{\text{LG}} \vert \Phi_{\text{exp}} \rangle|^2}$, where $\Phi_{exp}=\Phi_{\text{HyGG}}(w_0^{\text{exp}})$, $w_0^{\text{exp}}$ is the experimental beam waist that has been considered equal to $w_0^{\text{HyGG}}+\delta$, with $\delta=0.062$ mm, and $w_0^{\text{HyGG}}$ is the beam waist chosen for the computer-generated holograms. Such a quantity $\delta$ has been added to take into account experimental imperfections. 
    The different colors are associated to classes of states characterized by similar theoretical predictions. In particular, $S_{r,i}$ and $S_{c,i}$ are the balanced superposition of extremal walker states with real or complex coefficients, respectively, and $Q_i$ are the elements of the Fourier basis.}
    \label{fig:Rap}
\end{figure}

\section{Hologram-based measurement}
\label{ologramma}

Measurement and manipulation of OAM, and in general of spatial properties of the light, can be performed through holograms. These are diffraction gratings that allow to manipulate the phase and the amplitude of the impinging electromagnetic field \cite{arrizon2007pixelated,bolduc2013holo,heckenberg1992generation}.
Thanks to two arising devices for beam shaping, the SLM and digital micro-mirrors~\cite{rubinszteindunlop2016roadmap}, the employment of computer-generated holograms is increasing.
To measure an arbitrary OAM state, we inject in a  single-mode fiber the field whose transverse spatial profile has been, previously,  modulated by an hologram.
For instance, let $I$ be the field of the incident beam and $I_m$ be the field encoded in the holograms. The output state in the far field will be the convolution between the two modes $I*I_m$ \cite{mair2001entanglement,Qassim2014}.
Measuring the signal coupled in the fiber that selects the Gaussian component, we estimate the mutual overlap between the two fields.
Following this procedure, it is possible to define a set of orthonormal states of light, compute the holograms, and characterize the incident beam in this basis.
It is worth noting that such technique is equivalent to projective measurements in quantum mechanics and then can be employed both at classical and single-photon level.
 
 \begin{figure*}[tbh]
    \centering
    \includegraphics[width=1\textwidth]{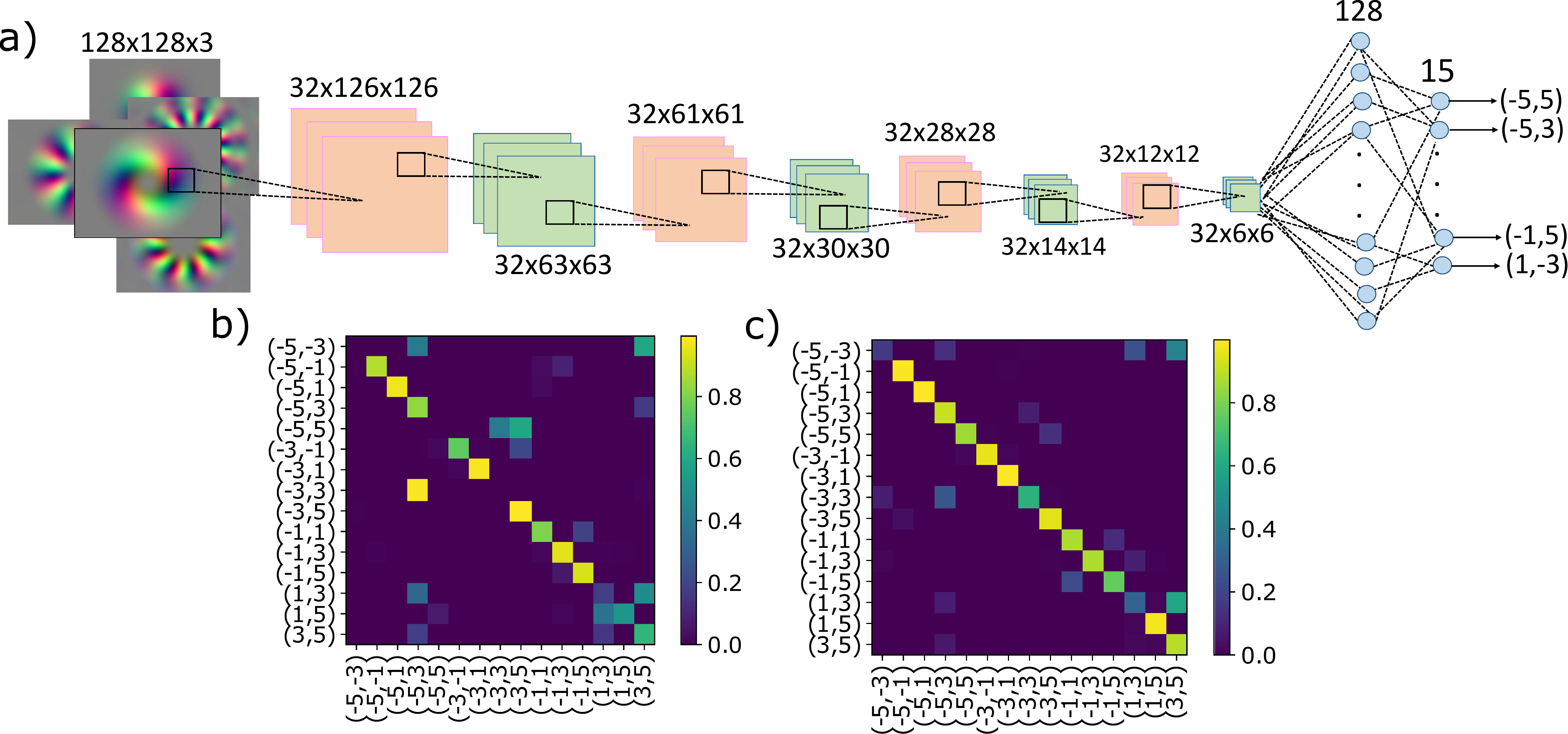}
    \caption{\textbf{Convolutional Neural Network.} a) Schematic representation of the CNN architecture. For each layer the correspondent dimension is shown in the figure. The Convolutional layers are represented in red, while the max pooling ones are shown in green. The last two layers are two fully connected ones. The specific classification task is performed by the last layer, i.e. the softmax one. b-c) Truth-tables for the two models: b) LG model and c) HyGG model. The matrix elements have been averaged over 100 experimental images, and the percentage of images belonging to the $i$-th class (row) classified by the CNN in the $j$-th class (column) are represented by the different colors.}
    \label{fig:Fid1}
\end{figure*}

To measure the output states produced by our experimental platform, we used a polarizing beam splitter to project the polarization state on the horizontal basis. Subsequently, the OAM detection is achieved by displaying the different elements of the orthogonal basis on a spatial light modulator (see Fig.~\ref{fig:app}). The accordance between the reconstructed state and the target one has been quantified exploiting the state distribution on the orthogonal basis.
We consider as figure of merit the \textit{fidelity} $\mathcal{F}$, that corresponds to square of the absolute value of overlap between observed and reference state.

In Fig.~\ref{fig:Fid} the cyan and purple bars report the values of $\mathcal{F}$ obtained with the LG and HyGG model, respectively, following the procedure for different classes of states that has been described above.  
We first set the platform parameters so as to generate OAM eigenstates, which corresponding to the computational basis in quantum mechanical language. One basis includes the target state as predicted by the LG model and the other by the HyGG one. 

To further assess the generation and measurement capabilities of our apparatus, we also showcase the generation of more complex OAM states. We focus in particular on different  balanced superpositions of the extremal walker positions, and four elements of the Fourier basis associated to the Hilbert space of the walker (QFT). The values are explicitly reported in the Appendix B. The average fidelities obtained with HyGG and LG models are $0.9831 \pm 0.0005$ and $0.9671 \pm 0.0010$, respectively. The higher fidelities achieved with the HyGG model confirms that this approximation describes more accurately the conditions faced in our experiments. 

 Importantly, we observe also higher coupling efficiencies in the single-mode fiber (Fig. \ref{fig:Rap}) for each hologram associated with an engineered target state. In particular, encoding a field $I_m$ in the hologram which projects the incoming beam onto the Gaussian one, an increase in the coupling efficiency corresponds to an higher mutual overlap between the two fields. To support this conclusion, in Fig. \ref{fig:Rap} we report the agreement between the measured coupling efficiency ratio ($\eta_{\text{HyGG}}/\eta_{\text{LG}})$ and the theoretical expectation $D$ calculated from the employed holograms and the actual experimental states. Such theoretical value is computed as the squared absolute value of the inner product between the target state used to generate the hologram in the HyGG model, and the experimental state 
 over the inner product between the target state, used to generate the hologram in the LG model and the experimental state. This result is a further proof of the 
 enhancement achieved in the measurement of the experimental states engineered through the QW platform. The capability of significantly improving the detection efficiency represents a fundamental aspect, especially at single photon level. Indeed, in this scenario photon losses undermine the security and the feasibility of quantum communication and cryptography protocols. Furthermore, when moving to multi-photon protocols, losses affect the amount of detected signal as $\eta^n$, being $n$ the number of involved photons, and thus an improvement in $\eta$ will result in a magnified overall efficiency.

\section{Machine-learning-based classification}
\label{ML}

Machine learning and neural-networks-based algorithms have been recently used, among other things, to classify structured light~\cite{giordani2020machine,doster2017machine,zhao2018mode} and for noise compensation~\cite{liu2019deep,li2018joint,Krenn13648, krenn_2014,Lohani_18, NEARY2020126058,Xie:15,lohani2018turbulence,turpin2018light,bhusal2020spatial}.
CNNs are a class of neural-network architectures especially suited to process images for classification and regression tasks. Its specific structure allows to recognize translation-invariant features, making CNNs effective to recognise images produced in realistic experimental conditions.

In this section, we apply CNNs to classify experimental images of VVBs using only \textit{simulated} ones in the training set. 
Remarkably, we show that improving the theoretical model used to generated the simulated images brings a significant improvement in the resulting accuracy.
We stress that this is not an obvious result, due to the simulated images remaining, in many regards, significantly different than experimental ones. The improvements in classification accuracy brought forward by the improvements in the simulation model indicate that the CNN appears to base its classification on physically significant features of the images.

In particular, we will compare the performances of CNNs trained with images generated using either the LG or the HyGG model. Notice however that, for the sake of the training, in both cases a validation set comprising only experimentally generated images is used. We have observed that the inclusion of the latter is necessary in order to achieve satisfactory performances in both cases.

Specifically, we use CNNs composed by four convolutional layers each of one is followed by a max-pooling layer~[cf. Fig.~\ref{fig:Fid1}].
On each convolutional layer we use $32$ filters of size $3\times3$ with ReLU activation function. For the pooling layers we apply the max operation to blocks of size $2\times2$. The final classification is performed by a fully connected layer with $128$ nodes followed by a softmax layer. For the training we used $400$ simulated images (either via the LG or the HyGG model) for each of the 15 involved classes. The classes are characterized by a fixed value for the $\theta$ parameter equal to $\pi/2$, $\beta \in [0,2\pi]$ and they can be distinguished through the different values of $m_1$ and $m_2$ which can assume each possible combination with $m_1 ,m_2 \in [-5,-3,-1,1,3,5]$. 
At each training step we use a mini-batch of $30$ simulated images to update the parameters, and a fixed validation set of $1500$ experimental images to assess the CNN's performance.

We then tested the classification accuracy of the network with $1500$ new experimental images. In Fig.~\ref{fig:Fid1} b) and c)  we report the best  performances where each VVB state has been labelled according to the OAM numbers of the two beams in the state superposition. 
These networks give average accuracies of $0.632 \pm 0.097$ and $0.815 \pm 0.065$ for LG and HyGG models, respectively.
The errors represent here the standard deviations on the average accuracy per class. We also report the accuracies obtained performing $22$ separate training instances, to provide information on how easily CNNs giving accurate results are obtained. We find the average accuracies to be $0.553 \pm 0.013$ (LG) and $0.662 \pm 0.019$ (HyGG), respectively (more details can be found in the accompanying Appendix C).

In addition to the post-processing on the training set that has been exploited in each trained network, we highlight the results obtained taking into account a physical and experimental issue related to the possibility of making an error in the rotation and/or calibration of the WPs. 
In order to account for this kind of error we generated a new set of theoretical images, both for the LG and the HyGG model, with a random error in the experimental WP angles up to a maximum value equal to $3\degree$. These images were added to the training set and we repeated the same analysis illustrated before, the accuracy increases for both models achieving a mean value of $0.605\pm0.018$ (LG) and  $0.765\pm0.015$ (HyGG) [cf. Appendix C].


In general, these results showcase how training a CCN with images generated using HyGG model allows us to achieve significantly higher classification accuracy than those resulting from the LG-trained case. 

\section{Conclusions}
\label{conc}
In this paper, we have experimentally demonstrated significantly enhanced and optimized  performances of detection techniques, by using a refined modeling of the incoming OAM modes.

In particular, we describe the states engineered through a QP-cascaded platform by using the HyGG modes. This allows us to obtain significant improvements in the OAM characterization process both in the classical and in the quantum regime. On one hand, we showed that the use of HyGG to evaluate computer-generated holograms allows to reach higher fidelities in quantum state discrimination. This is also accompanied by an increase in the coupling efficiency of each hologram. This feature represents a substantial improvement at the single-photon level, where the overall detection efficiency and accuracy represent a fundamental requirement for robust and secure quantum communication and quantum cryptography protocols. Furthermore, in light of the significant impact that losses have in the quantum regime, enhanced efficiencies are crucial in this context. On the other hand, we verified the enhancement reached in an example of learning that consists in the use of a CNN to classify experimental VVBs using only simulated images for the training set. Comparing the accuracy obtained using images simulated through the HyGG and the LG model, we can assert that the former captures more successfully the features of the experimental images. Our results are thus very promising for applications to various problems in quantum technologies that exploit information encoded in OAM states.

\section*{Acknowledgments}
We acknowledge support from the European Union's Horizon 2020 research and innovation programme (Future and Emerging Technologies) through projects CANCER SCAN (grant nr. 828978) and TEQ (grant nr. 766900), the DfE-SFI Investigator Programme (grant 15/IA/2864), COST Action CA15220, the Royal Society Wolfson Research Fellowship (RSWF\textbackslash R3\textbackslash183013), the Leverhulme Trust Research Project Grant (grant nr.~RGP-2018-266), the UK EPSRC (grant nr.~EP/T028106/1).

\appendix

\begin{table}[htp]
\centering
\begin{tabular}{ 
  |c
  | c
  |c | }
\hline
State & LG model & HyGG model \\
\hline
$\ket{-1}$ & \;\;$0.9832 \pm 0.0019$\;\;  & $0.9864 \pm 0.0013$ \\
\;&\;&\;\\
$\ket{1}$ &  \;\;$0.9806 \pm 0.0021$\;\; & $0.9857 \pm 0.0012$ \\
\;&\;&\;\\
$\ket{3}$ &  \;\;$0.9793 \pm 0.0036$\;\; & $0.9904 \pm 0.0011$ \\
\;&\;&\;\\
$\ket{-3}$&  \;\;$0.9686 \pm 0.0059$\;\; & $0.9761 \pm 0.0024$ \\
\;&\;&\;\\
$\ket{-5}$ &  \;\;$0.9876 \pm 0.0017$\;\; & $0.9900 \pm 0.0013$ \\
\;&\;&\;\\
$\ket{5}$ &  \;\;$0.9873 \pm 0.0018$\;\; & $0.9955 \pm 0.0007$ \\
\;&\;&\;\\
$\frac{1}{\sqrt{2}}(\ket{-5}+\ket{5})$ & \;\;$0.9675 \pm 0.0050$\;\;  & $0.9874 \pm 0.0021$ \\
\;&\;&\;\\
$\frac{1}{\sqrt{2}}(\ket{-5}-\ket{5})$ & \;\;$0.9829 \pm 0.0039$\;\;  & $0.9929 \pm 0.0006$ \\
\;&\;&\;\\
$\frac{1}{\sqrt{2}}(\ket{-5}-i\ket{5})$ & \;\;$0.9622 \pm 0.0038$\;\;  & $0.9768 \pm 0.0021$ \\
\;&\;&\;\\
$\frac{1}{\sqrt{2}}(\ket{-5}+i\ket{5})$ & \;\;$0.9780 \pm 0.0022$\;\;  & $0.9900 \pm 0.0015$ \\
\;&\;&\;\\
$\text{QFT}_{1}$ & \;\;$0.9620 \pm 0.0034$\;\;  & $0.9832 \pm 0.0024$ \\
\;&\;&\;\\
$\text{QFT}_{2}$ & \;\;$0.9308 \pm 0.0039$\;\;  & $0.9657 \pm 0.0027$ \\
\;&\;&\;\\
$\text{QFT}_{3}$ & \;\;$0.9165 \pm 0.0058$\;\;  & $0.9727 \pm 0.0029$ \\
\;&\;&\;\\
$\text{QFT}_{4}$ & \;\;$0.9525 \pm 0.0057$\;\;  & $0.9707 \pm 0.0037$ \\
\;&\;&\;\\
\hline
Average value & $0.9671 \pm 0.0010$ & $0.9831 \pm 0.0005$\\
\hline
\end{tabular}
\caption{ \textbf{Fidelity values.} The table shows the values of the fidelity for the engineered states. In the second and third column we report the values obtained with the holograms generated with the LG and the HyGG model, respectively.}
\label{Tab:Fidelity}
\end{table}

\begin{table*}[htp]
\centering
\begin{tabular}{ 
  |c
  | c
  |c 
  |c| c|}
\hline
State & $\eta_{\text{LG}}$ & $\eta_{\text{HyGG}}$ & $\eta_{\text{HyGG}}/\eta_{\text{LG}}$ & {$D$}\\
\hline
$\ket{-1}$ & \;\;$0.0951 \pm 0.0036$\;\;  & $0.1175 \pm 0.0044$ & $1.235 \pm 0.064$ & $1.096$\\
\;&\;&\;&\;&\;\\
$\ket{1}$ &  \;\;$0.0835 \pm 0.0031$\;\; & $0.0931 \pm 0.0034$ &$ 1.116 \pm 0.058$& $1.096$\\
\;&\;&\;&\;&\;\\
$\ket{3}$ &  \;\;$0.0387 \pm 0.0014$\;\; & $0.0770 \pm 0.0029$ &$1.99 \pm 0.10$ & $1.739$\\
\;&\;&\;&\;&\;\\
$\ket{-3}$&  \;\;$0.0448\pm 0.0017$\;\; & $0.0918 \pm 0.0034$ &$2.05 \pm 0.11$ & $1.739$\\
\;&\;&\;&\;&\;\\
$\ket{-5}$ &  \;\;$0.01873 \pm 0.00069$\;\; & $0.05397 \pm 0.0020$ &$2.88 \pm 0.15$ & $3.12$\\
\;&\;&\;&\;&\;\\
$\ket{5}$ &  \;\;$0.01609 \pm 0.00059$\;\; & $0.0492 \pm 0.0018$ & $3.06 \pm 0.16$ & $3.12$\\
\;&\;&\;&\;&\;\\
$\frac{1}{\sqrt{2}}(\ket{-5}+\ket{5})$ & \;\;$0.00445 \pm 0.00016$ & $0.01121 \pm 0.00041$ &$2.52 \pm 0.13$ & $3.12$\\
\;&\;&\;&\;&\;\\
$\frac{1}{\sqrt{2}}(\ket{-5}-\ket{5})$ & \;\;$0.00362 \pm 0.00013$ & $0.01104 \pm 0.00041$  &$3.05 \pm 0.16$& $3.12$\\
\;&\;&\;&\;&\;\\
$\frac{1}{\sqrt{2}}(\ket{-5}-i\ket{5})$ & \;\;$0.00613 \pm 0.00022$\;\;  & $0.01232 \pm 0.00045$  &$2.01 \pm 0.11$& $3.12$\\
\;&\;&\;&\;&\;\\
$\frac{1}{\sqrt{2}}(\ket{-5}+i\ket{5})$ & \;\;$0.00456 \pm 0.00017$\;\;  & $0.01091 \pm 0.00040$  &$2.38 \pm 0.13$& $3.12$\\
\;&\;&\;&\;&\;\\
$\text{QFT}_{1}$ & \;\;$0.01380 \pm 0.00051$\;\;  & $0.02642 \pm 0.00098$ &$1.91 \pm 0.10$& $2.138$\\
\;&\;&\;&\;&\;\\
$\text{QFT}_{2}$ & \;\;$0.01158 \pm 0.00043$\;\;  & $0.02210 \pm 0.00081$&$1.91 \pm 0.10$& $2.093$ \\
\;&\;&\;&\;&\;\\
$\text{QFT}_{3}$ & \;\;$0.01221 \pm 0.00045$\;\;  & $0.0276 \pm 0.0010$ &$2.26 \pm 0.12$& $2.066$\\
\;&\;&\;&\;&\;\\
$\text{QFT}_{4}$ & \;\;$0.02631 \pm 0.00097$\;\;  & $0.0368 \pm 0.0014$ &$1.400 \pm 0.073$& $2.317$ \\
\hline
\end{tabular}
\caption{\textbf{Coupling efficiency values.} In the table we report the coupling efficiencies for each hologram programmed both with the LG model ($\eta_{\text{LG}}$) and with the HyGG model ($\eta_{\text{HyGG}}$). In the third column we report the ratios between $\eta_{\text{HyGG}}$ and $\eta_{\text{LG}}$. These ratios are compared with the theoretically predicted quantity $D=\frac{|\langle \Phi_{\text{HyGG}} \vert \Phi_{\text{exp}} \rangle|^2}{|\langle \Phi_{\text{LG}} \vert \Phi_{\text{exp}} \rangle|^2}$, where $\Phi_{exp}=\Phi_{\text{HyGG}}(w_0^{\text{exp}})$ and $w_0^{\text{exp}}$ is the experimental beam waist. }
\label{Tab:Eff}
\end{table*}

\section{Theoretical Modelling details}
The experimental platform used to engineer arbitrary beams endowed with the Orbital Angular Momentum (OAM) degree of freedom, is implemented with a cascade of 5 $q$-plates inter-spaced by a set of waveplates. In the main text, we exploited a refined model that employs Hypergeometric-Gaussian modes to describe the beam propagation through this platform. In order to describe more clearly the formalization of the Hypergeometric model, we analyze a specific example concerning the generation of the vector vortex beam (VVB) described by $\theta = \pi /2$, $\beta = 0$, $m_{1}=-1$ and $m_{2}=1$. To generate this VVB, the parameters of each coin operator are set to act as the identity transformation in the polarization Hilbert space. Consequently, the input beam on the first $q$-plate is:
\begin{equation}
    E_{\text{in}} = \text{LG}_{0,0} \ket{H}.
\end{equation}
Solving the Fresnel's integral reported in Eq. \ref{Eq.Fresnel} for a topological charge $q=1/2$ and $\alpha_{0}=0$, the output beam from the first $q$-plate is:
\begin{equation}
    E_{I\text{step}} = \frac{e^{-i kz}}{\sqrt{2}}\left(\text{HyGG}_{-1,1}\ket{R}+\text{HyGG}_{-1,-1}\ket{L} \right).
\end{equation}
Thus, exploiting Eq. \ref{eq:HyGG_as_LG}, the beam is approximated as the superposition of Laguerre-Gaussian beam up to order $k=3$:
\begin{equation}\label{Eq:FirstStep}
    E_{I\text{step}} \approx \frac{e^{-i kz}}{\sqrt{2}}\left(\sum_{k=0}^{3}A_{-1,k}\text{LG}_{k,1}\ket{R}+\sum_{k=0}^{3}A_{-1,k}\text{LG}_{k,-1}\ket{L} \right).
\end{equation}
This mode is used as input in the second $q$-plate after propagation through a distance equal to $z_{2}$. By solving the Fresnel's integral for each of the Laguerre-Gaussian modes in \ref{Eq:FirstStep}, the output of the second step is found to be:
\begin{equation}
    E_{II\text{step}} = e^{-i k(z-z_{2})}\left(\sum_{p=0}^{3}C_{p}\text{HyGG}_{\;1+2p,0}\right)\frac{\ket{L}+\ket{R}}{\sqrt{2}},
\end{equation}
where the coefficients $C_{p}$ are obtained by grouping factors associated to Hypergeometric-Gaussian modes characterized by the same radial and azimuthal index. Therefore, by performing the same approximation of the first step we find:
\begin{equation}
    E_{II\text{step}} \approx e^{-i k(z-z_{2})}\left(\sum_{p=0}^{3}C_{p}\sum_{k=0}^{3}A_{1+2p,k}\text{LG}_{k,0}\right)\frac{\ket{L}+\ket{R}}{\sqrt{2}}.
\end{equation}
Iterating this procedure for the remaining steps, the output beam from the setup is:
\begin{eqnarray}
    & E_{\text{out}} = \frac{e^{-i k(z-z_{5})}}{\sqrt{2}} \Bigg(\sum_{p=0}^{3}C'_{2p-1}\text{HyGG}_{2p-1,1}\ket{R}+ \nonumber \\ & 
    +\sum_{p=0}^{3}C'_{2p-1}\text{HyGG}_{2p-1,-1}\ket{L} \Bigg)
\end{eqnarray}
where $z_{5}$ is the position of the fifth $q$-plate.

Consequently, using an equivalent argument it is possible to obtain the output description for each VVB and for an arbitrary OAM state. In particular, a general superposition of OAM states that involves azimuthal indexes $m$ ranging from $-5$ to $5$ is expressed as follow:
\begin{equation}
   E_{\text{out}} = e^{-i k(z-z_{5})} \sum_{m \in M} \sum_{p=0}^{4}C'_{2p-1}\text{HyGG}_{2p-1,m}
\end{equation}
where $M=\{-5,-3,-1,1,3,5\}$ and the coefficients $C'_{2p-1}$ also depend on the coin parameters used to engineer the target state.

\begin{table}[h]
\centering
\begin{tabular}{
  | c 
  | c
  | c | }
\hline
State ($m_{1}$, $m_{2}$) & LG model & HyGG model \\
\hline
(-5, -3) & 0  &  0.174\\
(-5, -1) & 0.879  & 0.993 \\
(-5, 1)  & 0.975    & 1 \\
(-5, 3)  & 0.832  & 0.915 \\
(-5, 5)  & 0  & 0.862 \\
(-3, -1) & 0.753  & 0.965  \\
(-3, 1)  & 0.984  & 0.998 \\
(-3, 3)  & 0  & 0.651 \\
(-3, 5)  & 0.995  & 0.961 \\
(-1, 1)  & 0.804     & 0.880 \\
(-1, 3)  & 0.949  & 0.878 \\
(-1, 5)  & 0.936  & 0.756 \\
(1, 3)   & 0.186  & 0.310 \\
(1, 5)   & 0.527     & 0.980 \\
(3, 5)   & 0.657  & 0.898 \\
\hline
Average value & $0.632 \pm 0.097$ & $0.815 \pm 0.065$\\
\hline
\end{tabular}
\caption{Prediction accuracies on the experimental images of the 15 classes under analysis obtained using the trained CNN. In the first column we report the OAM values of each class. In the second and third columns we show the fraction of correctly classified images for the two models. These latter values are the diagonal elements of the truth-tables shown in Fig. 4 b-c.}
\label{Tab:CNN}
\end{table}
 
 \section{Hologram-based state discrimination results}
In the following, we summarize the target states engineered at the single-photon level. Furthermore, we compare the fidelities obtained performing the measurement with the holograms generated with the LG model and with the HyGG model. As a first step, enhanced performances associated with the holographic technique have been verified for the elements of the computational basis $ \vert m \rangle=\{ \vert -5 \rangle, \vert-3 \rangle,\vert-1 \rangle,\vert 1 \rangle,\vert 3 \rangle,\vert 5 \rangle \}$. Subsequently, we have considered coherent superpositions of the extreme sites of the walker, both with real and complex coefficients: $\frac{\vert 5 \rangle + e^{i\beta}\vert -5 \rangle}{\sqrt{2}}$ where $\beta \in [0, \pi/2, \pi, 3\pi/2]$. Finally, to prove that these enhanced performances extends to general input states, we prepared four states of the Fourier basis associated to the Hilbert space of the walker, $ \frac{1}{\sqrt{6}}\sum_{j=1}^6 e^{\frac{i\pi j k}{3}}\vert j \rangle$ where $\vert j \rangle \in \{ \vert -5 \rangle, \vert-3 \rangle,\vert-1 \rangle,\vert 1 \rangle,\vert 3 \rangle,\vert 5 \rangle \}$ and $k=1,2,3,6$. In Table \ref{Tab:Fidelity}, we report the fidelities obtained for each target state with the LG and the HyGG model.
Moreover, in Table \ref{Tab:Eff} we show the coupling efficiency associated with each hologram, and the coupling improvement of the holograms programmed with the HyGG model compared to those obtained with the LG model.

\section{Convolutional Neural Network Results}

A CNN is a type of neural network specially designed to handle images and, analogously, spatially structured data. CNNs are composed of a sequence of layers of neurons, each one with a specific structure and fulfilling a specific purpose. The first layer applies filters on the images: a \textit{convolution} is performed between subsets of the image with a number of trained nodes referred to, in this context, as \textit{filters}. These filters are slid over the entire image in order to extract relevant features. Usually the following layers, as max- or average-pooling, down-sample the convoluted images and, in the case of classification, the extracted features are assigned to a class by a final fully-connected layer of neurons. This structure makes the CNN performances invariant for translation or misalignment in the images recording and, therefore, suitable in the experimental conditions. 

\begin{figure*}[!ht]
    \centering
    \includegraphics[width=1\textwidth]{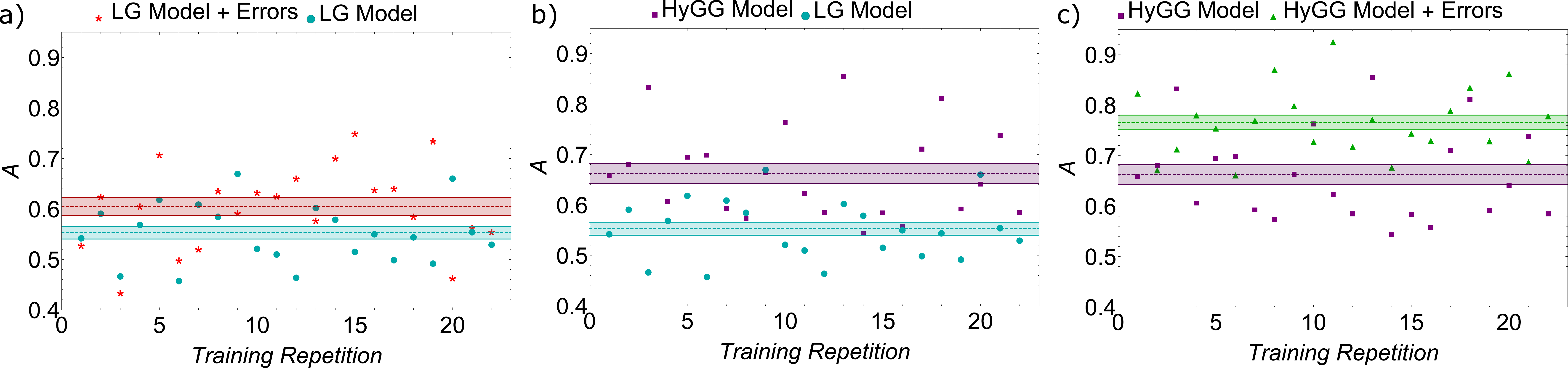}
    \caption{\textbf{ Training Accuracy.} In the image, we show the distinct values of accuracy reached for different and independent training. a) We compare the CNN performances trained with simulated images obtained through the LG model (sky-blue) and theoretical images computed with the LG model taking into account experiment errors regarding the orientation of the waveplates (red). b) CNN accuracies obtained with a theoretical training set based on the LG model (sky-blue) and on the HyGG model (purple).
    c) We compare the CNN performances trained with simulated images obtained through the HyGG model (purple) and theoretical images computed with the HyGG model taking into account experiment errors regarding the orientation of the waveplates (greed). The sky-blue, red, purple, and green areas are the respective average values with their standard deviations: $0.553 \pm 0.013$, $0.605\pm 0.018$, $0.662\pm 0.019$, $0.765 \pm 0.015$. }
    \label{fig:CCN Error}
\end{figure*} 

To build and train a CNN we used the Python library Keras~\cite{chollet2015keras}, with TensorFlow~\cite{tensorflow2015-whitepaper} as backend. We then show a more complete comparison between the results obtained by training the CNN with images simulated through the LG model or through the HyGG model. In particular, we report in Table \ref{Tab:CNN} the diagonal elements of the truth-tables presented in panels b and c of Fig. \ref{fig:Fid1}. Each element is computed over a set of experimental images of 100 elements per each class, and indicates the fraction of correctly predicted input images. As clearly shown, the HyGG model allows to reach a higher accuracy for almost every tested class. 

Moreover, in Fig. \ref{fig:CCN Error} we report the maximum accuracy obtained for different and independent training. This graph shows the higher average performance reached by CNN using images simulated with the HyGG model. Besides, these performances could be further improved taking into account experimental imperfections of the apparatus. In particular, panels a) and c)  report the accuracy values reached by simulating the images through the LG model (red) and the HyGG model (green) that considers a possible error in the setting of the waveplates' orientation.

\section*{References}
\bibliography{mrefs}
\bibliographystyle{iopart-num}

\end{document}